# Incommensurate host-guest structures in compressed elements: Hume-Rothery effects as origin

**V F Degtyareva**

Institute of Solid State Physics Russian Academy of Sciences, Chernogolovka, Russia

E-mail: degtyar@issp.ac.ru

**Abstract.** Discovery of the incommensurate structure in the element Ba under pressure 15 years ago was followed by findings of a series of similar structures in other compressed elements. Incommensurately modulated structures of the host-guest type consist of a tetragonal host structure and a guest structure. The guest structure forms chains of atoms embedded in the channels of host atoms so that the axial ratio of these subcells along the c axis is not rational. Two types of the host-guest structures have been found so far: with the host cells containing 8 atoms and 16 atoms; in these both types the guest cells contain 2 atoms. These crystal structures contain a non-integer number of atom in their unit cell: $tI11*$ in Bi, Sb, As, Ba, Sr, Sc and $tI19*$ in Na, K, Rb. We consider here a close structural relationship of these host-guest structures with the binary alloy phase $Au_3Cd_5$-$tI32$. This phase is related to the family of the Hume-Rothery phases that is stabilized by the Fermi sphere – Brillouin zone interaction. From similar considerations for alkali and alkaline-earth elements a necessary condition for structural stability emerges in which the valence electrons band overlaps with the upper core electrons and the valence electron count increases under compression.

## 1. Introduction

Recent high pressure experiments show, structure and properties of elements change dramatically under pressure (see review papers [1-3] and references therein). Periodic table of elements if considered in compressed state looks very different from what we are familiar with at ambient conditions. The elements on the right side of the Periodic table become close-packed metals and superconductors, like Si and Ge, whereas simple metallic structures such as bcc and fcc of the elements from the left side of the Periodic table (groups I - II) transform under pressure to open and complex structures. A remarkable example is discovery of the incommensurate host-guest structure in Ba [4] - the first case of the self-hosting incommensurate structure in the element.

Structures with the same 8-atom host framework have been found in Sr, Sc, As, Sb and Bi [1,3,5]. A different structure of the same type was in Rb and K. In these elements, the host structure has the same tetragonal $I4/mcm$ space group as in Ba, Sr, Sc and Bi, but is made up of 16 atoms. Recently same structure was found in Na [6,7]. The lighter group-II element Ca transformed to similar 8-atom host structure with the commensurate axial ratio of the host and guest cells $c_G/c_H$ [8].

In this paper we investigate possible causes that contribute to the formation of the complex crystal structures in elements under pressure, which have similar crystal structures and atomic volumes for the elements from the left and from the right of the Periodic table. We suggest the electronic cause for the formation of its crystal structures.

## 2. Theoretical background and method of analysis

The crystal structure of metallic phases is defined by two main energy contributions: electrostatic (Ewald) energy and the electron band structure term. The latter usually favours the formation of superlattices and distorted structures. The energy of valence electrons is decreased due to a formation of Brillouin planes with a wave vector q near the Fermi level $k_F$ and opening of the energy pseudogap on these planes if $q_{hkl} \approx 2k_F$ [9]. Within a nearly free-electron model the Fermi sphere radius is defined as $k_F = (3\pi^2 z/V)^{1/3}$, where z is the number of valence electrons per atom and V is the atomic volume. This effect, known as the Hume-Rothery mechanism (or electron concentration rule), was applied to account for the formation and stability of the intermetallic phases in binary simple metal systems like Cu-Zn, and then extended and widely used to explain the stability of complex phases in various systems, from elemental metals to intermetallics [10-16].

The stability of high-pressure phases in elements are analyzed using a computer program BRIZ [17] that has been developed to construct Brillouin zones or extended Brillouin-Jones zones (BZ) and to inscribe a Fermi sphere (FS) with the free-electron radius $k_F$. The resulting BZ polyhedron consists of numerous planes with relatively strong diffraction factor and accommodates well the FS. The volume of BZ's and Fermi spheres can be calculated within this program. The BZ filling by the electron states ($V_{FS}/V_{BZ}$) is estimated, which is important for the understanding of electronic properties and stability of the given phase. For a classical Hume-Rothery phase $Cu_5Zn_8$, the BZ filling by electron states is equal to 93%, and is around this number for many other phases stabilized by the Hume-Rothery mechanism.

Diffraction patterns of these phases have a group of strong reflections with their $q_{hkl}$ lying near $2k_F$ and the BZ planes corresponding to these $q_{hkl}$ form a polyhedron that is very close to the FS. The FS would intersect the BZ planes if its radius, $k_F$, is slightly larger then $q_{hkl}/2$. One should keep in mind that in reality the FS would usually be deformed due to the BZ-FS interaction and partially involved inside the BZ. The ratio $2k_F/q_{hkl}$, called a "truncation" factor, equal usually 1.01 – 1.05, is an important characteristic for matching to Hume-Rothery rules. Thus, with the BRIZ program one can obtain the qualitative picture and some quantitative characteristics on how a structure matches the criteria of the Hume-Rothery mechanism.

## 3. Results and discussion

*3.1. Incommensurate host-guest structure tI19* in alkali metals*

Potassium transforms at ~20 GPa from fcc to the host-guest *tI*19* structure [1] with a decrease in coordination number from CN=12 to CN=9-10 and with a significant drop in atomic volume (about 10%) as considered in Section *3.2*. The minimal interatomic distance for K-*tI*19* is 2.832 Å which is *smaller* than the double ionic radii $2r_i$ = 3.1Å (for CN=9) and 3.18Å (for CN=10) estimated by Shannon [18] This implies for potassium an increase in the number of valence electrons per atom in the first *post-fcc* phase – K-*tI*19*.

Similar incommensurate host-guest structure *tI*19* was found in Rb and Na with the difference in the guest subcell (K: C-face-centered, Rb: body-centered, Na: monoclinic) [1,6,7]. The other type of host-guest structure assigned as *tI*11* is formed under pressure in Ba, Sr, Sc, Bi, Sb and As (see reviews [1,3]) and recently found in Ca [8]. We found structural relationship between atomic arrangement in these two types of host-guest structures and a binary simple metal phase $Au_3Cd_5$-*tI*32 [19]. Phases $Au_3Cd_5$ and $Au_5Cd_8$ exist on the phase diagram Au-Cd in the same composition range (~62 at% Cd) differing only by temperature: $Au_3Cd_5$ transforms to $Au_5Cd_8$ above 370°C. The crystal structure of $Au_5Cd_8$ is of $Cu_5Zn_8$-*cI*52-type well known as a Hume-Rothery phase stabilized by a Fermi sphere – Brillouin zone interaction [9]. Therefore stability of $Au_3Cd_5$-*tI*32 can be considered within the Hume-Rothery mechanism.

In the $Au_3Cd_5$-*tI*32, the arrangement of Au and Cd atoms in the space group *I4/mcm* is shown in figure 1 for the layer at z = 0. Cd atoms in position 16k form octagon (eight-sided polygon) – square nets same as K-host atoms in *tI*19*. Au atoms in position 8h form square – triangle nets like Bi-host atoms in *tI*11*. Atoms Cd in 4b and Au in 4a relate to guest atoms in chains in K and Bi host-guest structures with increased interatomic distances where a guest subcell is formed with an incommensurate $c_G/c_H$ ratio and the number of atoms in the structure is reduced to non-integer values.

Interestingly, that relation between two types of host-guest structures was discussed in [20]. For the host-guest structures in Ba and Rb "a remarkable duality" was found when the representation of structure is considered in terms of atoms and localized electrons ("blobs"). The appearance of the interstitial electron blobs has been described as the formation of an *electride* in which the interstitial electrons are the anions. The Ba-IV and Rb-IV structures are therefore related "by reversing the roles of the ions and electron blobs".

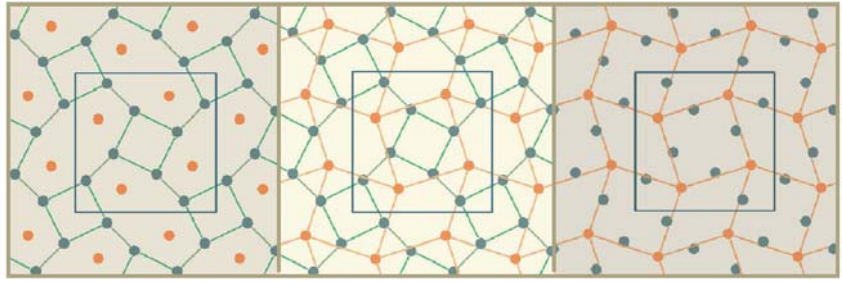

**Figure 1.** Atomic configuration for $Au_3Cd_5$-$tI32$, $I4/mcm$ in the layer at z=0 (middle). Cd atoms (green) in position 16k form octagon - square nets like K-host in $tI19*$ (left). Au atoms (red) in position 8h form square - triangle nets like Bi-host in $tI11*$ (right).

Axial ratio for $Au_3Cd_5$-$tI32$ is equal to c/a =0.499 and is comparable to $(c/a)_{host}$ for $tI19*$ equal to 0.485 for K at 20GPa, 0.491 for Na at 130GPa and 0.501 for Rb at 16.8GPa. For $tI11*$ in Bi $(c/a)_H$ = 0.489 also close to ~0.5, whereas in Ba and Sr $(c/a)_H$ is considerably higher (0.562 and 0.569, respectively). Therefore host-guest structures for Ba and Sr as well as for Ca are characterized in relation to another FS-BZ configuration. For our consideration we choose the structures $Au_3Cd_5$-$tI32$, K-$tI19*$ and Bi-$tI11*$. Characteristics of the discussed structures are given in table 1.

**Table 1.** Structure parameters of host-guest phases K-$tI19*$, Bi-$tI11*$ and $Au_3Cd_5$-$tI32$. The Fermi sphere radius $k_F$, the total number of Brillouin zone planes, the ratio of $2k_F$ to Brillouin zone vectors ($2k_F/q_{hkl}$) and the degree of filling of Brillouin zones by electron states $V_{FS}/V_{BZ}$ are calculated by the program BRIZ [17].

| Phase | K-(h-g)$tI19*$ | Bi-(h-g)$tI11*$ | $Au_3Cd_5$ |
|---|---|---|---|
| Structural data | | | |
| Pearson symbol | $tI19*$ | $tI11*$ | $tI32$ |
| Space group | $I4/mcm$(00γ)000s | $I'4/mcm$(00γ)0000 | $I4/mcm$ |
| P,T conditions | 22 GPa T=300K | 6.8 GPa T=300K | 0 GPa T=300K |
| Lattice parameters (Å) | $a_H$ = 9.767 $c_H$ = 4.732 $c_G$ = 2.952 γ= 1.603 | $a_H$ = 8.518 $c_H$ = 4.164 $c_G$ = 3.180 γ= 1.309 | a = 10.728 c = 5.352 c/a =0.499 |
| Number of atoms in cell | 19.2 | 10.62 | 32 |
| Atomic volume (Å³) | 23.50 | 28.326 | 19.25 |
| $V/V_0$ | 0.312 | 0.80 | |
| FS - BZ data from the BRIZ program | | | |
| z (number of valence electrons per atom) | 2.6 | 4.75 | 1.625 |
| $k_F$ (Å⁻¹) | 1.485 | 1.704 | 1.357 |
| Total number of BZ planes | 48 | 48 | 32 |
| Filling of BZ with electron states $V_{FS}/V_{BZ}$ (%) | 95 | 95 | 95 |

In our consideration we propose an overleap of upper core electrons with the valence band implying an increase in the number of valence electrons and involving the Hume-Rothery mechanism for understanding of complex phase formation. Figure 2 (upper) shows diffraction pattern and the constructed Brillouin zone with the planes close to the Fermi sphere counting the number of valence electrons per atom z = 1.625. The planes 420, 202 and 411 form a polyhedron with 32 planes accommodating well the Fermi sphere with the filling by electron states equal to 95% which meets the Hume-Rothery mechanism criteria.

To work out the number of valence electrons for K and Bi in their host-guest phases we use the following idea that the number of valence electrons in a Brillouin zone for the same Bravais lattice is held nearly constant: zN ≈ const, were N is the number of atoms in the cell, e.g. if the number of valence electrons increases, the number of atoms in the cell would decrease to keep the zN nearly constant. It was suggested by Jones (see ref. [9] p. 206) that "the large zone in k-space is defined by the Bravais lattice on which the structure is based and the Miller indices of its faces. Hence its volume and the number of states it contains, is independent of the number of atoms in the unit cell." To a first approximation, the Brillouin-Jones zone can be used to estimate the likely electron count for a participating element. For $Au_3Cd_5$ where z=1.625 and N=32 we have zN = 52. Taking the same Bravais lattice and the same Brillouin zone we obtain for K-$tI$19* with N=19.2 electron count z = (52/19.2) = 2.7 and for Bi-$tI$11* with N=10.6 electron count z = (52/10.6 ) = 4.9.

The valence electron number per atom for K and Bi given in table 1 is reduced to comply with degree of BZ filling ~ 95% and accepted to be 2.6 for K-$tI$19* and 4.75 for Bi-$tI$11*. Non-integer electron-per-atom counts appear due to overlap of $s$ (or $sp$) and $d$ levels and within the FS-BZ model only $s$ (or $sp$) electrons are taken into account as structure controlling electrons.

For Bi-$tI$11* host atoms (8h) are assumed to have 5 valence electrons per atom and guest atoms (there are 2.6 guest atoms in the host unit cell) have 4 electrons per atom giving the average electron count 4.75 per atom. The decrease in z from 5 to 4 for guest Bi atoms is assumed in the above consideration and is possible because a lowering of the upper empty $d$-level may occur with increasing temperature or pressure which would lead to the level hybridization. Following the same arguments, one can rationalize the appearance in Bi of the $oC$16 structure with the stability range around z=4 electrons per atom which was found to appear in Bi at same pressures as $tI$11* but at higher temperature. Stability of the $oC$16 type structure in elements and alloy compounds was

Similar, for K-(h-g) $tI$19* one can expect different valency for the host and guest atoms with more participation of $d$-electrons for guest atoms that provides strong bonding in chains of guest atoms with short distances and gives rise to effect of chain melting and different guest-atom-chains arrangements.

Diffraction patterns for host-guest structures K-$tI$19* and Bi-$tI$11* are shown in figure 2 with indication of $2k_F$ position for corresponding z. The BZ polyhedrons are constructed for commensurate approximants assuming γ equal 8/5=1.6 and 4/3=1.333, respectively. BZ's for h-g structures contain additional planes from guest reflections compared to BZ for $Au_3Cd_5$ that reduce slightly the volume of BZ leading to nearly full BZ. Therefore electron counts were defined from the 95% filling of the zone resulting in z=2.6 for K and 4.75 for Bi. The total number of BZ planes in the contact with the FS increases to 48 (see table 1) that should result on the substantial changes of physical properties at the transition to these host-guest structures.

Both $tI$11* and $tI$19* incommensurate host-guest structures form additional planes at the Fermi sphere compared to a commensurate basic structure leading to the gain in the electron band structure energy even if this is not favourable for the electrostatic term that prefers more symmetrical arrangement. This shift of balance in the structural energy may account for incommensurate structures of other types. For example, found in phosphorus above 100 GPa, P-IV, an orthorhombic structure with an incommensurately modulation wave vector along the c-axis was characterized by appearance of satellite reflections and hence by a formation of additional Brillouin zone planes near the FS [21]. It should be noted that the polyhedron formed by the first strongest reflection (211) is placed commonly inside the FS contacting with it by vertices (figure 2, middle) that reduces the band structure energy. Such effects were shown to account tetragonal fcc distortions in In alloys [22].

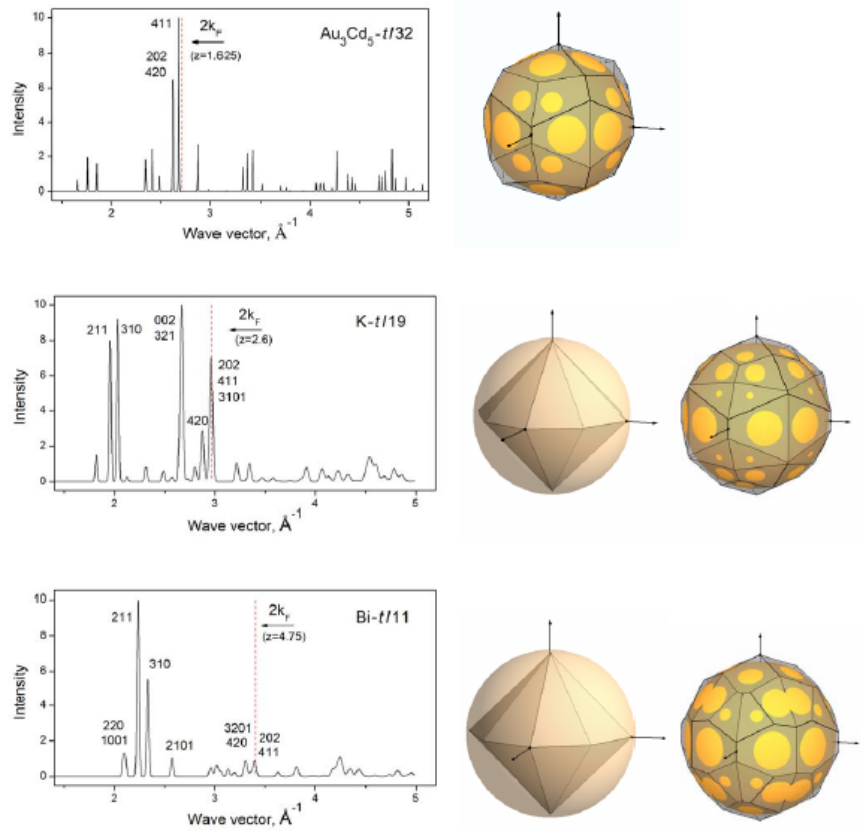

**Figure 2.** Calculated diffraction patterns (left) and corresponding Brillouin zones with the inscribed FS (right). The Brillouin zone of (211) planes is circumscribed by FS (middle). In the top panel – for $Au_3Cd_5$-$tI$32, in the middle panel – for K- $tI$19* and in the bottom panel – for Bi- $tI$11* with structural parameters given in table 1. The position of $2k_F$ for given z and the *hkl* indices of the principal planes are indicated on the diffraction patterns.

*3.2. Valence electron count in compressed alkali elements*

Alkali elements display under pressure very large compressibility. Change of interatomic distances along the pressure for Na and K are shown in figure 3 with indication of the coordination numbers (CN). Structures followed *fcc* with CN=12 are characterized by the decrease in coordination numbers and substantial decreases in the interatomic distances. Horizontal line on figure 3 indicates double ionic radius for CN=9 as suggested for Na and K by Shannon [18]. These values are 1.25Å × 2 = 2.5Å and 1.55Å × 2 = 3.1Å for Na and K, respectively. The interatomic distances for *post-fcc* structures lay below these values implying the core overlap at such compression. For structure stability it is necessary to assume core ionization – overlap the upper core electrons with the valence band electrons. Resulting valence electron count should be increased and univalent alkali metals turn into polyvalent metals.

Similar behaviour is expected for alkaline-earth elements as considered below for the host-guest structure in Ca. Changes in the valence electron counts for Ca from 2 to 3 - 3.5 are expected in the *post-fcc-bcc* structure with the simple cubic cell at pressure 32 GPa as suggested in [13].

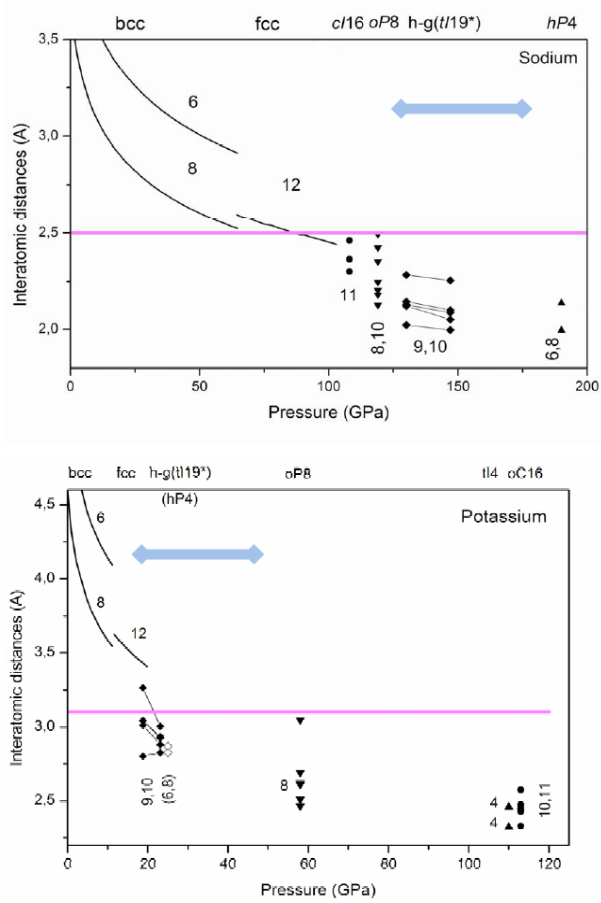

**Figure 3.** Interatomic distances in Na (upper panel) and K (lower panel) in the experimentally observed phases as a function of pressure (data taken from review paper [3]). The numbers show how many of those distances are present for one particular atom. Only distances from the first coordination shell are shown. The pressure regions for the h-g(*tI*19*) phase are indicated by arrows. Horizontal line indicates double ionic radius for CN=9 known for Na and K [18].

*3.3. Ca-VII with the commensurate host-guest structure*

Structural behaviour of Ca under compression in the *post-bcc* region differs essentially from that of heavier alkali-earth metals Sr and Ba. Only at pressure above 210 GPa Ca-VII adopts the host-guest structure [8] which is similar to host-guest structures found in Ba and Sr [1]. The latter two phases are incommensurate since host and guest subcells in the common tetragonal cell have irrational ratio of *c* parameters. Thus, $c_H/c_G$ was found ~1.388 for Ba-IV (at 12 GPa) and ~1.404 for Sr-V (at 56 GPa). The Ca-VII structure may be considered as having a similar host-guest cell with a commensurate ratio 4/3. Evaluated axial ratio for the basic *tI*32 cell is c/a=1.667, and one obtains by dividing to 3 for subcell $(c/a)_H = 0.556$. This value is close to $(c/a)_H$ for Ba and Sr (0.562 and 0.569, respectively). These differences define another choice of the Brillouin zone configuration than in the case of host-guest structures with $(c/a)_H \sim$ 0.5 for host-guest structures in alkali metals and group V elements, considered above.

The structure solution for Ca-VII was suggested in two models as *I4/mcm*-32 and *P4$_2$/ncm*-128, the latter is a supercell of the former with dimensions 2×2 ×1 [8]. Another example of a complex host-guest structure was reported recently for Ba-IVc (at 19 GPa) with $3\sqrt{2} \times 4\sqrt{2} \times 3$ supercell of the basic host-guest cell that results in 768 atoms in the unit cell [23]. The host-guest ratio in this structure has a commensurate value of 4/3 as in the case of Ca-VII. The basic structure for Ca-VII has tetragonal unit cell with 32 atoms, space group *I4/mcm*, and this structure has a prototype structure In$_5$Bi$_3$-*tI*32 as found in the Pauling File Database [24].

Structural relations of Ca-VII and In$_5$Bi$_3$ give opportunity to imply common features of electron structures. Figure 4 shows diffraction pattern for the basic *tI*32 structure for Ca and BZ configurations with accepting z~3.6 electrons per atom that satisfies well the Hume-Rothery mechanism. BZ filling by electron states is ~90%, as discussed in [13]. Similar conclusions can be suggested for the incommensurate host-guest structures of Ba and Sr. Assuming for Ba z~3.7 one obtains nearly same FS-BZ configuration as is shown in figure 4 for Ca-VII with the only difference that planes (314) for Ca will correspond to satellite guest planes (3101) type.

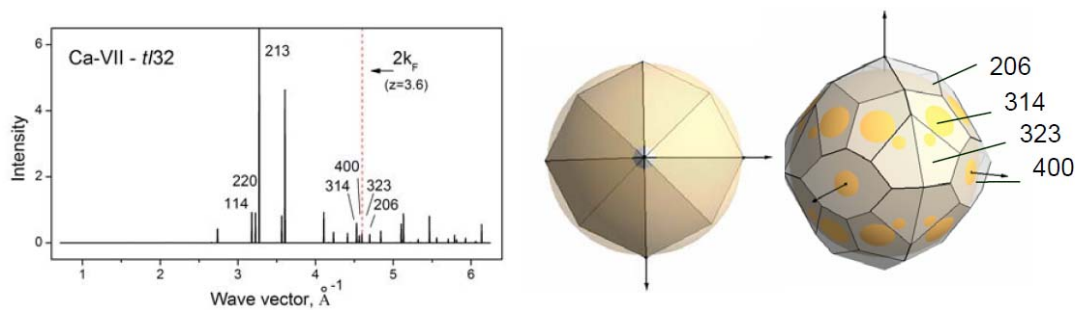

**Figure 4.** Calculated diffraction pattern of Ca-VII, *tI*32, *I*4/*mcm*, P = 241 GPa, *a* = 5.5099, *c* = 9.1825 Å (left); principal diffraction peaks *hkl* are assigned; the position of 2k$_F$ is indicated for z = 3.6 valence electrons by a vertical dashed line. The Brillouin zone of (213) planes is circumscribed by FS (middle); view along *c**. The Brillouin zone of group planes close to 2k$_F$ and the inscribed FS (right). The BZ–FS configuration satisfies the Hume–Rothery conditions.

## 4. Summary

Complex structures observed under pressure in elements of groups I and II as well as in group V with the incommensurate host-guest cell are analyzed to define main factors of unusual asymmetric atomic arrangements. The energy contribution of valence electron band structure increases under compression leading to formation of low-symmetry complex structures satisfying the Hume-Rothery rules. Two types of the host-guest structures have been found: with 8-atom and 16-atom host cells: *tI*11* in Ba, Sr, Sc, Bi, Sb, As and *tI*19* in Na, K, Rb. We consider here a close structural relationship of these host-guest structures with the binary alloy phase Au$_3$Cd$_5$-*tI*32. The latter phase is related to the family of the Hume-Rothery phases that is stabilized by the Fermi sphere – Brillouin zone interaction where a decrease in the electronic band structure energy occurs due to the contact of the Fermi sphere and Brillouin zone planes. An important characteristic is degree of Brillouin zones filling by electron states which depends on the count of valence electrons per atom and the number of atoms in the cell.

From these considerations for alkali and alkaline-earth elements a necessary condition for structural stability emerges in which the valence electrons band overlaps with the upper core electrons and the valence electron count increases under compression. Consideration of the core – valence band electron transfer may promote a better understanding of non-traditional behaviour of alkali and alkali-earth elements under significant compression.

**Acknowledgments**
The author thanks Dr Olga Degtyareva for valuable discussions. This work was partially supported by the Program of the Russian Academy of Sciences "The Matter under High Energy Density".